\documentclass[aps,prx,reprint,superscriptaddress,floatfix,longbibliography]{revtex4-2}

\usepackage{amsmath}
\usepackage{amssymb}
\usepackage{graphicx}
\usepackage{tabularx}
\usepackage{makecell}
\usepackage{bm}
\usepackage{gensymb}
\usepackage{mathbbol}
\usepackage{mathrsfs}
\usepackage{xcolor}
\usepackage[colorlinks=true,linkcolor=blue,citecolor=blue,urlcolor=magenta]{hyperref}
\usepackage{placeins}
\usepackage{comment}

\newcommand{\sumloop}{\mathop{\sum \kern-1.3em \circlearrowleft}}

\DeclareMathAlphabet{\mathpzc}{OT1}{pzc}{m}{it}

\AtBeginDocument{}
\AtBeginDocument{}
\AtBeginDocument{}

\newcommand{\braket}[1]{\ensuremath{\langle{#1}\rangle}}
\newcommand{\bra}[1]{\langle #1 |}
\newcommand{\ket}[1]{| #1 \rangle}

\newcommand{\iu}{\mathrm{i}}

\newcommand{\xzpfi}[1]{x_{\text{zpf},#1}}

\DeclareMathOperator{\diag}{diag}

\newcommand{\mdrvdet}{\Delta_\text{m}}
\newcommand{\Jc}{J_\text{c}}

\newcommand{\subidc}[1]{\textbf{(#1)}}

{\left(\smallmatrix}
{\endsmallmatrix\right)}

\newcommand{\temptext}[1]{\textcolor{orange}{#1}}
\begin{document}

\title{Programmable Synthetic Magnetism and Chiral Edge States in Nano-Optomechanical Quantum Hall Networks}

\author{Jesse J. Slim}
\affiliation{Center for Nanophotonics, AMOLF, Science Park 104, 1098 XG Amsterdam, The Netherlands}
\affiliation{Australian Research Council Centre of Excellence for Engineered Quantum Systems (EQUS),
School of Mathematics and Physics, University of Queensland, St Lucia, Queensland 4072, Australia}
\author{Javier del Pino}
\affiliation{Department of Physics, University of Konstanz, 78457 Konstanz, Germany}
\author{Ewold Verhagen}
\email{verhagen@amolf.nl}
\affiliation{Center for Nanophotonics, AMOLF, Science Park 104, 1098 XG Amsterdam, The Netherlands}

\date{\today}

%

\begin{abstract}
Artificial magnetic fields break time-reversal symmetry in engineered materials --- also known as metamaterials, enabling robust, topological transport of neutral excitations, much like electronic conduction edge channels in the integer quantum Hall effect. We experimentally demonstrate the emergence of quantum-Hall-like chiral edge states in optomechanical resonator networks. Synthetic magnetic fields for phononic excitations are induced through laser drives, while cavity optomechanical control allows full reconfigurability of the effective metamaterial response of the networks, including programming of magnetic fluxes in multiple resonator plaquettes. By tuning the interplay between network connectivity and magnetic fields, we demonstrate both flux-sensitive and flux-insensitive localized mechanical states. Scaling up the system creates spectral features that are precursors to Hofstadter butterfly spectra. Site-resolved spectroscopy reveals edge-bulk separation, with stationary phononic distributions signaling chiral edge modes. We directly probe those edge modes in transport measurements to demonstrate a unidirectional acoustic channel. This work unlocks new ways of controlling topological phononic phases at the nanoscale with applications in noise management and information processing.
\end{abstract}

\maketitle

\section{Introduction}

The discovery of the integer quantum Hall effect (IQHE) in a two-dimensional electron gas under a magnetic field~\cite{Klitzing1980new} sparked significant interest in new phases of matter with unique properties. As formalized by Thouless et al.~\cite{Thouless1982quantized}, the quantized Hall conductance in the IQHE is connected to a nontrivial topology of the bulk wavefunctions in momentum space. It emerges as the magnetic field breaks time-reversal ($\mathcal{T}$) symmetry, so that the electron’s wavefunction acquires a nonreciprocal Aharonov-Bohm (AB) phase~\cite{Cohen2019geometric} when traversing the magnetic vector potential. 
Interference from waves following multiple paths then creates an insulating bulk and chiral (one-way) conducting edge states, robust against backscattering and protected by the bulk's topology rather than microscopic details~\cite{Thouless1982quantized,Hasan2010colloquium}. 

As the IQHE is essentially a wave phenomenon~\cite{Laughlin1981quantized}, it was realized that it equally applies to classical waves~\cite{Haldane2008possible,Raghu2008}. The unique properties of the resulting \emph{bosonic} topological phases of matter incited considerable fundamental and technological interest in photonic and phononic settings~\cite{Ozawa2019topological,Shah2024}. However, to induce AB phases for such neutral excitations 
requires complex engineering to break $\mathcal{T}$-symmetry and mimic the effect of a magnetic field, without relying on charge. For electromagnetic waves, this is for example achieved with magneto-optical materials~\cite{Umucalilar2011,Wang2009observation,Klembt2018excitonpolariton} and for low-frequency sound through rotating fluids or coupled gyroscopes~\cite{Fleury2014sound,Nash2015topological}. The materials and system limitations involved in these approaches make scaling to the nanophotonic and nanomechanical domain a significant challenge. Moreover, they generally lack \emph{in situ} active tunability.

\begin{figure}[tb]
		\centering
		\includegraphics{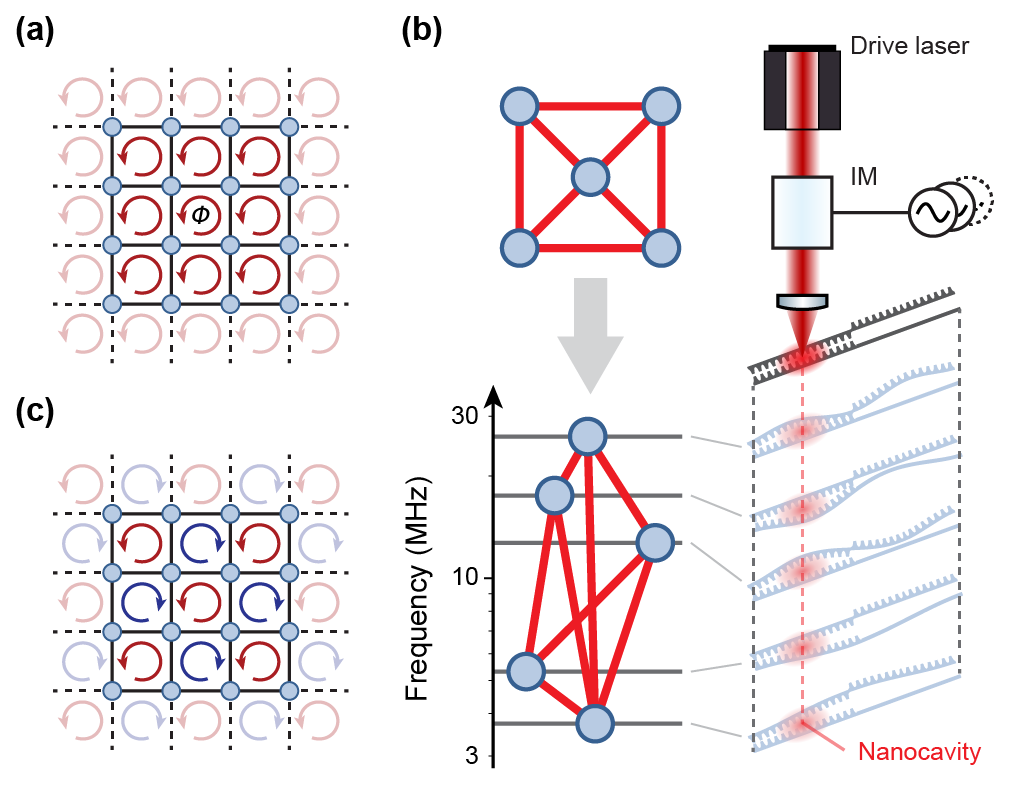}  
		\caption{\textbf{Constructing resonator networks with controllable synthetic magnetic fields.} 
		\subidc{a} Square lattice threaded by a uniform magnetic flux $\Phi$ per plaquette, inducing counterclockwise chirality (arrow). A small section is highlighted for clarity. 
		\subidc{b} Graph with 5 resonators connected by 8 hopping interactions. Resonators represent mechanical overtones of a nano-optomechanical cavity, which span a synthetic frequency dimension. Each interaction is generated by modulating the optical drive at a mechanical difference frequency, establishing full connectivity by superposing (incommensurate) modulation tones.
		\subidc{c} Control over each coupling phase allows for creating lattices with highly inhomogeneous magnetic fields, difficult to 
        achieve in natural materials.
		\label{mpq:fig:concept}
		}
\end{figure}

Dynamical modulation offers a powerful alternative to break $\mathcal{T}$ symmetry. Using time-varying parametric drives, gauge fields for light and sound can be engineered -- and reconfigured -- at will~\cite{Koch2010timereversalsymmetry,Nunnenkamp2011synthetic,Dalibard2011colloquium,Fang2012photonic,Sounas2017nonreciprocal,Darabi2020reconfigurable}. 
In these Floquet systems, harmonic modulation of the coupling between bosonic resonators with distinct resonance frequencies enables frequency conversion, inducing a hopping (beam-splitter) interaction between them. 
The modulation phase is imprinted on the transferred excitations, and the process is nonreciprocal: exchanging the initial and target resonators results in a phase pickup of opposite sign, analogous to the Peierls phase for an electron traveling in a magnetic vector potential. After connecting resonators in a loop, such nonreciprocal hopping phases allow the creation of an AB loop phase that corresponds to a $\mathcal{T}$-breaking \emph{synthetic} magnetic flux piercing the loop. In a single loop, AB interference enables controllable nonreciprocal transmission. 
In extended lattices, combining many Aharonov-Bohm loops across all unit cells, such as in Fig.~\ref{mpq:fig:concept}\subidc{a}, mimics a homogeneous magnetic field and enables synthetic quantum Hall phases~\cite{Koch2010timereversalsymmetry,Aidelsburger2011,Aidelsburger2018,Fang2012realizing,Goldman2016topological,Ningyuan2015time,Darabi2020reconfigurable}.

Implementing these ideas in the nanomechanical domain has so far remained elusive. Multimode cavity optomechanical systems constitute a promising route to do so, naturally providing time-varying potentials for light and sound that can induce frequency-converting nonreciprocal couplings and provide $\mathcal{T}$-breaking gauge fields%
~\cite{Ruesink2016nonreciprocity,Fang2017generalized,Bernier2017nonreciprocal,Peterson2017demonstration,Verhagen2017optomechanical,Xu2019nonreciprocal,Mathew2020synthetic}. 
It was envisioned that topological phases for sound and light could be implemented in optomechanical lattices, specifically Chern insulators akin to the IQHE~\cite{Peano2015topological,Schmidt2015optomechanical}. While those first proposals relied on the resolved-sideband limit, requiring precise tuning of many high-$Q$ photonic resonators, recently, time-modulated lower-$Q$ cavities have offered a flexible alternative to imprint gauge fields for MHz frequency nanomechanical resonators~\cite{Mathew2020synthetic,delPino2022nonhermitian,Wanjura2023quadrature,Slim2024optomechanical}. While optomechanical couplings have been used to map ($\mathcal{T}$-symmetric) mechanical topological states using optical modes~\cite{Ren2022} and microwave topological states using mechanical resonators~\cite{Youssefi2022}, the realization of a Chern insulator that breaks $\mathcal{T}$ through optomechanical driving has remained an outstanding challenge.

In this work, we bring quantum Hall physics into the nanomechanical domain using a versatile optomechanical platform that exploits multiple nanomechanical modes in a `synthetic dimension'. Optical driving fully controls the coupling strength \emph{and} phase between each pair of resonators. The coherent hopping phase is nonreciprocal, unlike a trivial reciprocal phase that could be established through retardation. This control enables the creation of arbitrary resonator networks with broken $\mathcal{T}$ symmetry that feature multiple AB loops with individually tunable synthetic magnetic fluxes. We implement increasingly complex networks to study the role of artificial magnetic fluxes in multi-plaquette systems. We find that the relative handedness of neighboring plaquettes significantly impacts dynamics, and controls the symmetries and localization of mechanical states within the network. Finally, in a four-plaquette lattice, we launch and track a chiral edge mode that witnesses the emergence of the IQHE in a minimal nanomechanical network. This provides an essential advance in the pursuit to realize topological phases of sound induced by light~\cite{Peano2015topological}, and opens the ability to study the physical and technological implications of such phases of matter.

\section{Programmable network Hamiltonians with Aharonov-Bohm phases}

The employed device, described in more detail in~\cite{delPino2022nonhermitian,Slim2024optomechanical}, consists of a sliced photonic crystal nanobeam~\cite{Leijssen2015strong} and is shown in Fig.~\ref{mpq:fig:concept}\subidc{b}. 
It supports multiple non-degenerate mechanical modes with frequencies $\Omega_j$ ranging from $3.7$~MHz to $26$~MHz, each dispersively coupled to the optical field of a telecom-wavelength nanocavity (linewidth $\kappa/2\pi = 320$~GHz). The optical resonance frequency shifts by $\sum g_{0,j} x_j$, where $g_{0,j}$ is the vacuum optomechanical coupling rate of mode $j$ and $x_j$ its displacement in units of its zero-point motion amplitude $x_{\mathrm{zpf},j}$. Operating in the unresolved sideband regime ($\Omega_j \ll \kappa$), the cavity enables optical readout of both thermal and driven mechanical motion, as displacements $x_j$ modulate the reflected intensity of a detuned probe laser. This reveals five optically active mechanical resonances in the thermomechanical spectrum, with linewidths $\gamma_j/2\pi \approx 1-7$~kHz and coupling rates $g_{0,j}/2\pi\approx 2-6$~MHz (see Table~\ref{tab:device_params}). 

The cavity is also illuminated with a drive laser detuned from the cavity resonance frequency by $\Delta = \kappa/2\sqrt{3}$ to induce a strong optomechanical spring effect through radiation pressure back-action. Modulating its intensity at the frequency difference $\Omega_j - \Omega_k$ between  modes (referred from now on as resonators) $j$ and $k$ creates an effective mechanical beamsplitter interaction, at a rate $J_{jk} = c^m_{jk}\bar{n}_c \sqrt{3 g_{0,j} g_{0,k}} / 2\kappa=J_{kj}$, where $\bar{n}_c$ and $c^m_{jk}$ is the average cavity photon population and modulation depth, respectively. 
Crucially, each modulation tone's phase offset imprints a phase shift $\varphi_{jk}$ on the interaction, akin to the Peierls phase imprinted by a magnetic vector potential on electron hopping~\cite{Mathew2020synthetic}. 

As all mechanical frequency differences in the system are incommensurate, we can establish multiple mechanical couplings simultaneously by superposing the necessary modulation tones on the single detuned drive laser. The resulting temporal modulation of the optomechanical spring effect thus allows programming the effective Hamiltonian of a resonator network through modulation depths and phases of the various modulation tones.
The Hamiltonian for $N$ resonators, in terms of the resonator annihilation operators $a_j$ in frames rotating at $\Omega_j$, then reads
\begin{align}\label{eq:Hamiltonian}
H=\sum_{j,k=1,k\neq j}^{N,N}J_{jk}e^{-\iu\varphi_{jk}} a_j^\dag a_k, &&\varphi_{kj}=-\varphi_{jk}.
\end{align}
Hamiltonian~\eqref{eq:Hamiltonian} conserves phonon number and exhibits a $U(1)$ gauge symmetry, where gauge transformations $a_j \mapsto a_j e^{i\delta_j}$ correspond to shifts in the time origin that modify the coupling phases $\varphi_{jk}$. However, the presence of closed loops (`plaquettes') in the system crucially allows for AB interference,
giving rise to a finite, gauge-invariant flux $\Phi_p = \sumloop{}_{jk \in p} \varphi_{jk}$ for each plaquette  $p$.
In addition to the coherent dynamics governed by Eq.~\eqref{eq:Hamiltonian}, each resonator is coupled to an independent thermal bath with an occupation of $n_j \sim 10^6$, leading to thermal fluctuations and dissipation at a rate $\gamma_j$.  We routinely achieve coupling strengths $J_{jk}$ that exceed dissipation, i.e., $J_{jk} > \gamma_j,\gamma_k$, bringing the system into the strong mechanical coupling regime. 

Similar to the adjacency matrix that encodes network connectivity~\cite{chartrand2012introductory} or the impedance matrix that governs current flow in circuits~\cite{kemmerly1993engineering}, the matrix $\mathbf{J}$ that encodes the complex elements $J_{jk}e^{-\iu \varphi_{jk}}$ in Eq.~\eqref{eq:Hamiltonian} captures the strength and phase of all couplings between resonators. Each non-zero entry in $\mathbf{J}$ thus defines the mechanical network topology. In the experiment, adjusting the amplitudes and phases of the modulation tones allows independent control over each matrix element, while deterministic phase relations are obtained by referencing all tones to a common clock (see~\cite{delPino2022nonhermitian}). This control makes it possible to engineer $\mathbf{J}$ to mimic complex coupling geometries (e.g., lattices or complete graphs) threaded by synthetic fluxes of arbitrary complexity, including highly inhomogeneous magnetic fields as exemplified in Fig.~\ref{mpq:fig:concept}c. 

The simplest closed-loop network is a single plaquette, coupled by equal rates $J_{jk} = J$ and threaded by a flux $\Phi = \varphi_{12} + \varphi_{23} +\cdots+ \varphi_{N1}$. Here, AB interference shifts the eigenfrequencies,
\begin{align}
    \epsilon_k = 2J \cos\left(\frac{2\pi k + \Phi}{N}\right),
    \label{eq:AB_loop_freqs}
\end{align}
of its momentum eigenmodes, labelled by $k$,
\begin{equation}
    \tilde{a}_k = \frac{1}{\sqrt{N}}\sum_{j=1}^{N} e^{-\iu 2\pi kj/N}a_j,
\label{eq:AB_loop_states}
\end{equation}
expressed in the gauge where $\Phi$ is evenly distributed across links ($\varphi_{12}=\varphi_{23}=\cdots=\varphi_{N1}=\Phi/N$). Any flux $\Phi \neq 0,\pi$ breaks $\mathcal{T}$-symmetry, lifting the degeneracy between the eigenmodes $\tilde{a}_k$ --- which exhibit a chiral phase advance between resonators of $2\pi k/N$. In prior experiments with $N=3$, where $k = \{-1, 0, 1\}$, we observed that interference between the $\tilde{a}_k$ states produced clockwise (counterclockwise) chiral Rabi oscillations for $\Phi =\pi/2$ ($\Phi =-\pi/2$) in the time evolution~\cite{delPino2022nonhermitian}. 

\begin{figure}[t]
		\centering
		\includegraphics{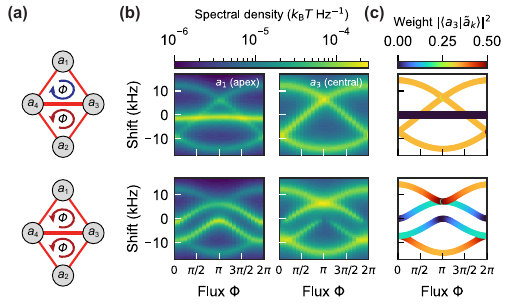}  
		\caption{\textbf{Interference between two adjacent plaquettes in a diamond configuration.} 
		\subidc{a} Four resonators $a_j$ form a diamond configuration with two adjacent plaquettes, pierced by opposite fluxes $\Phi$, $-\Phi$ (top) or equal fluxes $\Phi$ (bottom). The perimeter coupling rate is $J/(2\pi) = 5$~kHz, and the central link is coupled at $\Jc = J\sqrt{2}$. \subidc{b} Measured thermomechanical spectra around the frequency of the apex resonator $a_1$ (left) and central resonator $a_3$ (right) for varying flux $\Phi$. The spectra for the other apex and central resonators are similar. In the diamond with opposite fluxes (top), eigenmodes with flux-independent localization emerge. One antisymmetric mode, $\tilde{a}_\text{apex}$, is localized at the apex and remains unaffected by $\Phi$, while the other modes are delocalized and tune like those of a three-mode plaquette with $\Jc = J\sqrt{2}$ (Appendix~\ref{sec:apx:diamond_analysis}). For equal-handed fluxes (bottom), all eigenmodes tune with flux, both in frequency and localization. The central link couples opposite momentum states, and at $\Phi = 0$ and $\Phi = \pi$, their superpositions localize entirely on either the apex or central resonators. \subidc{c} Weight of central resonator $a_3$ in each hybridized eigenmode $\tilde{a}_k$ of the diamond.
		\label{mpq:fig:diamond-spectra}
		}
	\end{figure}

\section{Interference with multiple magnetic flux plaquettes}

Crucially, the collective phenomena in larger lattices pierced by magnetic fluxes originate from the interference of excitations on \emph{multiple} plaquettes in the lattice. For example, in the Harper-Hofstadter model~\cite{Hofstadter1976} interference between adjacent plaquettes leads to the insulating bulk and localization of chiral states on the edge of a network that are characteristic of the IQHE. We study such emergent behavior by constructing nano-optomechanical networks containing multiple flux plaquettes, leveraging the platform's reconfigurability. While flux tunes the eigenfrequencies $\epsilon_k$ (Eq.~\eqref{eq:AB_loop_freqs}) of a single AB loop with equal couplings $J$, its eigenstates remain evenly distributed for any $\Phi$ ($\langle a_j^{\dagger}a_j\rangle \equiv$ cnst.). But as soon as a network contains more than one loop, interference between adjacent plaquettes leads to eigenstates with inhomogeneous distributions, controlled by flux. The four-mode `diamond' network shown in Fig.~\ref{mpq:fig:diamond-spectra}a features two AB loops $a_4$-$a_1$-$a_3$ and $a_3$-$a_2$-$a_4$, fused along the central link $a_3$-$a_4$ and pierced by independent fluxes. The relative handedness of the fluxes critically impacts the diamond's spectrum and the eigenstate localization. 

With opposing flux chiralities $\Phi, -\Phi$, the phase vorticities for each loop align along the central link, and the net flux through the perimeter vanishes. We thus choose a gauge where both fluxes are sustained by the central link to simplify calculations. For equal perimeter couplings $J$, destructive interference then leads to an antisymmetric mode $\tilde{a}_\text{apex} = (a_1 - a_2)/\sqrt{2}$, localized solely on the apex resonators $a_1, a_2$. The localization keeps $\tilde{a}_\text{apex}$  unaffected by the central coupling rate $J_c$ and frequency-insensitive to $\Phi$. In Fig.~\ref{mpq:fig:diamond-spectra}b, the thermomechanical spectra (top) show mode $\tilde{a}_\text{apex}$ as a flat band near zero detuning, exclusive to the apex resonators’ sidebands. In this system, the apex localization, representing a compact localized state—where amplitude is confined to a few select sites—stems purely from geometry and remains robust under varying flux~\cite{Aoki1996}. However, the same coupling principle allows engineering nanomechanical $0$- and $\pi$-flux rhombic chains in our platform, where localization arises from both geometry and Aharonov-Bohm interference, leading to ``Aharonov-Bohm caging.’’
~\cite{Vidal1998aharonovbohm, Leykam2018artificial, Mukherjee2018experimental, Tang2020flat,Kremer2020}. 

The remaining three eigenmodes depend on both $\Jc$ and $\Phi$, affecting their frequency and localization. However, when $\Jc = J\sqrt{2}$, all eigenstates become flux-independent, and are given by $\tilde{a}_\text{apex}$ and
\begin{equation}
    \tilde{a}_k = \left[ \left(a_1 + a_2\right) / \sqrt{2} + e^{-\iu 2\pi k/3} a_3 + e^{\iu 2\pi k/3} a_4 \right]/\sqrt{3},
\end{equation}
with $k=\{-1, 0, 1\}$. Interestingly, $\tilde{a}_k$ preserve the phase-chiral character of the two three-mode loops forming the diamond, with eigenfrequencies tuning exactly as those of a single three-mode loop, as shown in Fig.~\ref{mpq:fig:diamond-spectra} (top) and detailed in Appendix~\ref{sec:apx:diamond_analysis}.

As Fig.~\ref{mpq:fig:diamond-spectra} (bottom) shows, the situation is strikingly different when both plaquettes are threaded by equally handed fluxes $\Phi$, $\Phi$. Their vorticities along the central link are antiparallel, so that multi-loop interference causes \emph{all} eigenstates to have flux-dependent weights. The perimeter is pierced by a net flux of $2\Phi$, which we can distribute evenly over its links. In the basis of eigenstates $\tilde{a}_k$ of perimeter momentum $k=\{-1, 0, 1, 2\}$, given in Eq.~\eqref{eq:AB_loop_states}, 
the hopping matrix -- order $\{\tilde{a}_0, \tilde{a}_2, \tilde{a}_1, \tilde{a}_{-1}\}$ -- becomes block-diagonal
\begin{equation}
\mathbf{J} = \begin{bmatrix}
\epsilon_0 + \Jc/2 & -\Jc/2 & 0 & 0\\
-\Jc/2 & \epsilon_2 + \Jc/2 & 0 & 0\\
0 & 0 & \epsilon_1 - \Jc/2 & \Jc/2 \\
0 & 0 & \Jc/2 & \epsilon_{-1} - \Jc/2
\end{bmatrix},
\end{equation}
where $\epsilon_k$ are given in Eq.~\eqref{eq:AB_loop_freqs} for flux $\Phi \to 2\Phi$. The central coupling $\Jc$ mixes the opposite momentum states $\tilde{a}_0 \leftrightarrow \tilde{a}_2$ and $\tilde{a}_1 \leftrightarrow \tilde{a}_{-1}$, which attain degenerate $\epsilon_k$ for $\Phi = 0, \pi$, respectively. Consequently, a spectral gap opens at those fluxes of width $J_c$ (see Appendix~\ref{sec:apx:diamond_analysis} and Fig.~\ref{fig:apx:opposed_diamond_tuning}), where (anti)symmetric superpositions of opposite momentum states are formed. These localize excitations in the apex or central resonators, as observed in Fig.~\ref{mpq:fig:diamond-spectra}b and c, bottom, for the central site $a_3$ by disappearing sidebands at $\Phi=0, \pi$, while the disappearance of apex sidebands is masked at $\Jc/J = \sqrt{2}$. Finally, at $\mathcal{T}$-breaking fluxes $\Phi \neq 0,\pi$, the $\epsilon_k$-degeneracies lift and all modes delocalize. Evidently, the equal-handed diamond no longer mimics its constituent loops.

\begin{figure}[t]
		\centering
		\includegraphics{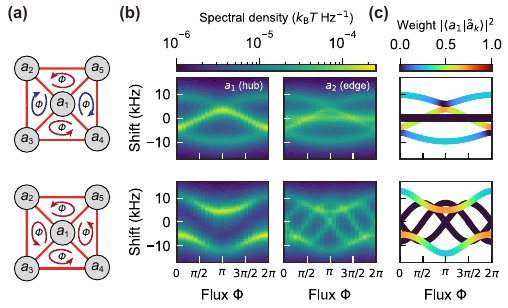}
		\caption{\textbf{Multi-plaquette interference in a wheel graph.}
		\subidc{a} Hub resonator $a_1$ is coupled to four perimeter resonators $a_j$ in a wheel graph configuration with equal rates $J=J_\text{s}$. The wheel has four plaquettes pierced by either alternating fluxes $\Phi$, $-\Phi$ (top, $J/(2\pi) = 3$~kHz) or equal-handed fluxes $\Phi$ (bottom, $J/(2\pi) = 4$~kHz).
		\subidc{b} Thermomechanical spectrum of the hub $a_1$ (left) and perimeter $a_2$ (right) for varying flux $\Phi$. With alternating fluxes, the spectrum mirrors the opposed-flux diamond, forming two degenerate, flux-insensitive eigenmodes that decouple from the hub, while the other three modes delocalize over all resonators. With equal fluxes, two modes delocalize across all resonators, and three phase-chiral modes localize on the perimeter. \subidc{c} Weight of hub resonator $a_1$ in the wheel's hybridized eigenmodes $\tilde{a}_k$.
 		 \label{mpq:fig:wheel5-spectra}}
\end{figure}

\section{Edge localization and integer quantum Hall physics}

When moving to larger lattices with magnetic flux, interference between plaquettes plays an essential role in the emergence of the features associated with quantum Hall phases. In infinite periodic lattices, such as the square-lattice Harper-Hofstadter model, a constant magnetic field controls the fractal complexity in the spectrum known as the Hofstadter butterfly, related to the competition between the lattice length scale and the magnetic length~\cite{Hofstadter1976}. A larger lattice uncovers more of the butterfly's details by accommodating a broader range of commensurate and incommensurate ratios, see Appendix~\ref{sec:apx:W5}. In a finite Hofstadter system, the characteristic chiral edge modes appear in the gaps of the butterfly spectrum, with energy localized on the outer edge of the lattice while the magnetic flux renders the bulk insulating. Similar effects are observed in other planar discretizations, including triangular, hexagonal Kagome lattices, and even hyperbolic tilings~\cite{Gumbs1997,avron2014study,agazzi2014colored,Du2018,Stegmaier2022}. 

\begin{figure*}[t!]
		\centering
		\includegraphics{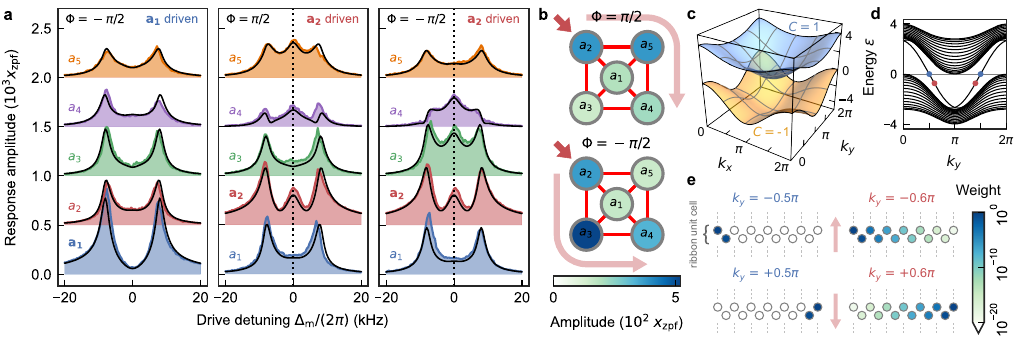}
		\caption{\textbf{Transport along the perimeter of a wheel graph.} 
		\subidc{a}~Amplitude response of the five-mode wheel network with equal-handed $\Phi=\pm\pi/2$ (Fig.~\ref{mpq:fig:wheel5-spectra}, bottom) to continuous wave driving of hub resonator $a_1$ (left) and perimeter resonator $a_2$ (middle and right). Coupling rates $J/(2\pi) = J_\text{s}/(2\pi) = 4$~kHz. Driving perimeter resonator $a_2$ excites the edge mode at zero detuning, while driving hub resonator $a_1$ does not. Measured amplitudes agree with the predicted responses (black lines, Appendix~\ref{sec:apx:w5_response_calcs}), with all parameters (coupling $J$, dissipation $\gamma_j$ and driving strength $f_j$) determined independently.
		\subidc{b}~\mbox{Amplitude response} at each site when resonantly driving $a_2$ ($\mdrvdet=0$, dotted lines in a). Clockwise (counter-clockwise) chiral transport along the perimeter is observed for $\Phi=\pi/2$ ($\Phi = -\pi/2$) as vibration transfer competes with decay. Differences in clockwise/counter-clockwise transport stem from disorder in dissipation rates (Table~\ref{tab:device_params}).
        \subidc{c}~Two-dimensional band structure of a Union Jack (UJ) lattice pierced by flux $\Phi=\pi/2$. Two bands are separated by a gap, with opposite Chern numbers $C = \pm 1$.
        \subidc{d}~Band structure of a UJ ribbon, finite in the $x$-direction (width: $33$ sites) and infinite in $y$. Two edge states traverse the topological gap, localized on either boundary of the ribbon. 
        \subidc{e}~Transverse edge state profiles in a unit cell of the UJ ribbon, at different values of $k_y$ (colored dots in panel \textbf{d}). States are localized at one edge of the ribbon, depending on propagation direction. The transverse decay length depends on $k_y$. 
        For $k_y = \pm \pi/2$, edge states fully localize on the outer two sites. 
		\label{mpq:fig:wheel5-transport}
		}
\end{figure*}

The mechanisms at play can be observed in systems that contain even just five resonators. We construct the five-mode wheel graph \(W_5\) (Fig.~\ref{mpq:fig:wheel5-spectra}\subidc{a}), with four resonators \(a_j\) forming a ring (rates \(J\)) and connected to a central hub resonator \(a_1\) (rates \(J_\text{s}\)). With oppositely handed fluxes through adjacent plaquettes (Fig.~\ref{mpq:fig:wheel5-spectra}, top), the vorticities of neighbouring plaquettes in $W_5$ align along their common links. This results in flux-insensitive antisymmetric eigenmodes,  $(a_2 - a_4)/\sqrt{2}$ and $(a_3 - a_5)/\sqrt{2}$, similar to the diamond’s apex state. For equal couplings \(J_\text{s} = J\), the remaining states behave like a three-mode loop with unequal couplings $J,J\sqrt{2}$ (Fig.~\ref{mpq:fig:wheel5-spectra}\subidc{b}, top). Appendix~\ref{sec:apx:W5} presents further details. Interestingly, at the special ratio $J_\text{s}/J = \sqrt{2}$, all states become flux-independent and delocalized, and the system acts as a fusion of its constituent loops. 

A striking shift occurs with equally handed fluxes, where the thermomechanical spectrum reveals states confined to the perimeter and tunable with flux (Fig.~\ref{mpq:fig:wheel5-spectra}, bottom). We will call these states `edge states' because of their confinement to the perimeter of this small lattice. Writing the Hamiltonian in the momentum basis of the edge states $b_k$, in a gauge where the spokes carry no phase yields
\begin{align}
    H_{W_5,\circlearrowright} = J_s \left( a_1^\dagger b_0 + b_0^\dagger a_1 \right) + \sum_{k=-1}^2 \epsilon_k b_k^\dagger b_k,
\end{align}
where $\epsilon_k$ are given by Eq.~\eqref{eq:AB_loop_freqs} for $N=4$ and $\Phi \to 4\Phi$ (perimeter net flux). The spoke couplings $J_s$ hybridize the hub state $a_1$ with the zero-momentum edge state $b_0$, opening a spectral gap as observed in Fig.~\ref{mpq:fig:wheel5-spectra}b and c, bottom. At $\Phi = \pm\pi/2$, only the edge state $b_{\pm 1}$ remains at zero detuning.

To study the character of the zero-detuning edge state, we probe continuous wave transport at $\mathcal{T}$-breaking $\Phi = \pm \pi/2$ (Fig.~\ref{mpq:fig:wheel5-transport}). Driving the hub resonator $a_1$ at \(\mdrvdet = 0\) fails to excite the edge state, but driving the perimeter resonator $a_2$ elicits a sharp response near \(\mdrvdet = 0\) across all perimeter resonators, with the hub's response remaining flat. Crucially, as coherent energy transfer through the network competes with dissipation, the relative heights of the zero-detuning peaks reveal chiral transport along the perimeter (Fig.~\ref{mpq:fig:wheel5-transport}), with energy flowing clockwise or counterclockwise depending on the sign of the flux. We thus observe chiral transport along the perimeter of the $W_5$ network, enabled by AB interference between plaquettes. As the single hub site $a_1$ is shared among the four plaquettes that make up $W_5$, destructive interference suppresses the excitation of $a_1$ as the edge wave propagates. These mechanisms ensure that the basic characteristics of the chiral edge states that exist in Chern insulators featuring the integer quantum Hall effect can be recognized in optomechanical networks, even if they are composed of a modest number of resonators. As such, these experiments witness the emergence of the IQHE in small optomechanical lattices, including edge confinement, chiral transport, and destructive interference in the bulk.

\section{Discussion and conclusions}

The connection of the above observations in the $W_5$ network to the IQHE in extended systems can be made more concrete in multiple ways. Indeed, the $W_5$ lattice could be extended to a larger lattice with topological band structure by continuously adding resonator nodes. For example, a square tiling of the $W_5$ graph forms a so-called `Union Jack (UJ)' lattice, a.k.a. tetrakis square tilling.~\cite{Stephenson1970,grunbaum1987tilings}. This lattice forms a Chern insulator, featuring bands with opposite Chern numbers $C=\pm1$ separated by a direct band gap, shown in Fig.~\ref{mpq:fig:wheel5-transport}(c). A detailed analysis is presented in Appendix~\ref{sec:apx:w5_topology}. Boundaries of the UJ lattice under a uniform flux support chiral edge states that span the gap, as illustrated in the dispersion of a finite-width ribbon in Fig.~\ref{mpq:fig:wheel5-transport}(d). The edge states, depicted in Fig.~\ref{mpq:fig:wheel5-transport}(e) for various different wavevectors, are localized near the edge, with the degree of localization determined by the spectral position of the state's energy within the gap. 
Alternatively, the $W_5$ lattice could be extended to a triangular or square lattice with equal flux along each plaquette, which are also known Chern insulators with full topological band gaps.

A natural question is how the optomechanical platform we exploit can be expanded to employ larger numbers of modes, and thus more complex networks. First, we recognize that a larger number of physically separated beams could be used, as envisioned for example in~\cite{Mathew2020synthetic}. This would expand the system by adding a spatial lattice dimension to the synthetic (frequency) mode dimension that we use here, in combination with multiple optical cavities. To keep the number of control tones manageable, some of the mechanical modes in the network could have equal frequencies, such that single modulation tones can address multiple connections in the network. Indeed, only a finite set of independently tunable hopping phases is needed to implement the IQHE, due to gauge invariance~\cite{Fang2012realizing}. Second, we envision that even in a single optical cavity, many more mechanical modes could be coupled than we have demonstrated in this work. The physical phenomena we demonstrated here require coupling rates $|J_{jk}|$ to significantly exceed the mechanical linewidths $\gamma_j,\gamma_k$, i.e., strong mechanical coupling and normal mode splitting. Since $|J|$ is of the order of the maximal optical spring shift, the above requirement is equivalent to having the optomechanical cooperativity $C=4g_0^2\bar{n}_\mathrm{c}/(\kappa\gamma)$ significantly exceed unity. This can be reached for a larger amount of mechanical modes by for example reducing the optical linewidth $\kappa$, or by reducing the mechanical linewidth $\gamma$. The mechanical linewidth in the platform studied here would be reduced naturally by more than an order of magnitude by operating at cryogenic temperatures~\cite{Leijssen2017nonlinear}. But many different optomechanical resonator system across various scales and materials reach the regime of large cooperativity~\cite{Aspelmeyer2014cavity}, and one can thus imagine the use of optomechanical lattices in systems that naturally feature many mechanical modes, such as phononic hole arrays~\cite{Halg2022} in a cavity, phononic waveguides coupled to an optical nanocavity~\cite{Patel2018singlemode}, multiple levitated particles coupled to a joint cavity field~\cite{Rieser2022}, or high-frequency bulk-acoustic wave resonators~\cite{Kharel2019}.

In conclusion, we employed in this work a versatile nano-optomechanical platform to engineer phononic networks threaded by artificial magnetic fields, demonstrating Aharonov-Bohm interference between multiple loops. We found that the relative handedness of neighboring loops greatly impacts their dynamics: Networks with loops that have their vorticity aligned along shared links (i.e. staggered fluxes in contiguous plaquettes) behave like their constituent loops, while those with opposing vorticities show richer, flux-dependent, localization. Specifically, in the $W_5$ network, we uncovered a chiral edge mode that can be associated with topologically protected edge states in lattices with broken $\mathcal{T}$-symmetry. This minimal demonstration of a nanomechanical IQHE marks an essential step in the realization of Chern insulators for sound at the nanoscale, and complements other experiments that achieve helical, bidirectional edge modes without breaking time-reversal symmetry~\cite{Ren2022}.

The demonstrated optomechanical Aharonov-Bohm control could enable nanoscale devices with slow sound or phononic localization, enhancing energy-efficient control and acoustic sensing through flat band engineering and enhanced sound-matter interactions
~\cite{Yang2015a,Ma2021}. Furthermore, this platform's weak, tunable nonlinearities from optomechanical interactions, combined with controllable connectivity, offer potential for analog computing, including Ising optimizers~\cite{Mahboob2016electromechanical} and neuromorphic computers~\cite{Markovic2020}.
Finally, the proposed principles could extend to hybrid quantum nanomechanical systems~\cite{Chu2017quantum}, where
qubit-induced strong nonlinearity and flat bands enable strongly correlated chiral states akin to those in superconducting cavity arrays~\cite{Roushan2017chiral,Rosen2024}.

\section*{Acknowledgments}
This work is part of the research programme of the Netherlands Organisation for Scientific Research. It is funded by the European Union, supported by ERC Grants 759644 (TOPP) and 101088055 (Q-MEME). Views and opinions expressed are however those of the authors only and do not necessarily reflect those of the European Union or the European Research Council Executive Agency. Neither the European Union nor the granting authority can be held responsible for them. 
J.J.S. acknowledges support from the Australian Research Council Centre of Excellence for Engineered Quantum Systems (EQUS, CE170100009). J.d.P. acknowledges financial support from the ETH Fellowship program (Grant No. 20-2 FEL-66).

\FloatBarrier

\clearpage

\appendix

\section{Experimental methods}\label{sec:apx:exp_methods}
The experimental set-up and methods for fabricating and characterising the nano-optomechanical device, calibrating the programmable interactions, phase referencing, read-out and driving of the resonator networks are described in detail in refs.~\cite{delPino2022nonhermitian,Wanjura2023quadrature,Slim2024optomechanical}. Table~\ref{tab:device_params} lists the relevant parameters of the five nanomechanical modes and the optical mode used in this study, determined as described in~\cite{delPino2022nonhermitian}. 

\begin{table}[h]
    \centering
    \begin{tabular}{c|ccc}
            & $\Omega_j/2\pi$ &       $g_{0,j}/2\pi$ & $\gamma_j/2\pi$ \\
    Mechanical mode    & (MHz)             & (MHz)     & (kHz) \\
    \hline
    $a_1$       & $\phantom{0}3.7$  & $5.3$       & $1.2$ \\
    $a_2$       & $\phantom{0}5.3$  & $5.9$       & $2.5$ \\
    $a_3$       & $12.8$            & $3.3$       & $2.7$ \\
    $a_4$       & $17.6$            & $3.1$       & $4.4$ \\
    $a_5$       & $26.3$            & $1.9$       & $6.9$ \\
    \hline
            & $\omega_0/2\pi$   &             & $\kappa/2\pi$ \\
    Optical mode & (THz)             &             & (GHz)    \\
    \hline
    $c$ & $195.451$         &             & $320$
    \end{tabular}
    \caption{Characteristics of the five mechanical modes $j$ and single optical mode used in this study. All mechanical resonances are supported by the on-chip nano-optomechanical device and coupled to a common photonic crystal optical nano-cavity. 
    The mechanical resonance frequencies $\Omega_j$, vacuum optomechanical coupling rates $g_{0,j}$ and damping rates $\gamma_j$ are listed, as well as the frequency $\omega_0$ and linewidth $\kappa$ of the optical resonance. Mechanical dissipation rates $\gamma_j$ include a contribution from photothermal backaction~\cite{delPino2022nonhermitian}.}
    \label{tab:device_params}
\end{table}

\section{Analysis of the diamond network}
\label{sec:apx:diamond_analysis}
In this section, we include further analysis of the diamond network shown in Fig.~\ref{mpq:fig:diamond-spectra}. First, we show that the dynamics of the chirality-opposed diamond (which has aligned vorticity along the central link), dubbed $\uparrow\downarrow,\Diamond$ in Fig.~\ref{mpq:fig:diamond-spectra}a, top, can be mapped onto the three-mode AB loop. The coupling matrix of a three-mode loop with one phase-carrying link of strength $J_a$ and two reciprocal links of equal strength $J_b$ is given by
\begin{equation}
    \mathbf{J}_3 = \begin{bmatrix}
    0                   & J_a e^{-\iu \Phi}  & J_b \\
    J_a e^{\iu \Phi}   & 0                 & J_b \\
    J_b                 & J_b               & 0 \\
    \end{bmatrix},
\end{equation}
where $\Phi$ is the flux that pierces the loop. Next, we express the  matrix $\mathbf{J}_{\uparrow\downarrow,\Diamond}$ of the chirality-opposed diamond in the basis $\{\tilde{a}_\text{apex}, a_2, a_3, \tilde{a}_\text{sym}\}$, where $\tilde{a}_\text{sym} = (a_1+a_2)/\sqrt{2}$ is the symmetric apex mode, to find 
\begin{equation}
    \mathbf{J}_{\uparrow\downarrow,\Diamond}' = \begin{bmatrix}
    0   & 0                   & 0                   & 0 \\
    0   & 0                   & J_c e^{-\iu \Phi}    & J\sqrt{2} \\
    0   & J_c e^{\iu \Phi}   & 0                   & J\sqrt{2} \\
    0   & J\sqrt{2}                   & J\sqrt{2}                   & 0 \\
    \end{bmatrix} = \begin{bmatrix}
        0 & \\
          & \mathbf{J}_3
    \end{bmatrix}.
    \label{eq:apx:aligned_diamond_ham_mapped}
\end{equation}
From Eq.~\eqref{eq:apx:aligned_diamond_ham_mapped}, we see that the antisymmetric apex mode $\tilde{a}_\text{apex}$ is decoupled, while the remaining modes $\{a_2, a_3, \tilde{a}_\text{sym}\}$ evolve as a three-mode AB loop with couplings $J_a \to J_c$, $J_b \to J \sqrt{2}$. Moreover, when $J_c = J \sqrt{2}$, the effective three-mode loop exhibits equal couplings and is rotationally invariant.

\begin{figure}[tb]
		\centering
		\includegraphics{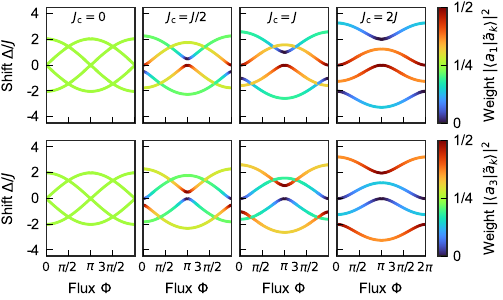}
		\caption{\textbf{Momentum-state interactions in the equal-handed diamond}. Calculated eigenfrequency spectra of a diamond network pierced by fluxes $\Phi$ of equal handedness (cf. Fig.~\ref{mpq:fig:diamond-spectra}a, bottom) for increasing central link strength $J_c$. Colors indicate the weight of the apex (top row) or central (bottom row) resonators in each eigenmode. Turning on $J_c > 0$ couples perimeter states of opposite momenta, opening a spectral gap as these become degenerate for $\Phi = 0,\pi$. The resulting symmetric and antisymmetric superpositions fully localize either in the apex or central resonators.
        \label{fig:apx:opposed_diamond_tuning}
        }
\end{figure}

For the equal-handed ––vorticity-opposed–– diamond shown in Fig.~\ref{mpq:fig:diamond-spectra}a, bottom, we calculate the spectra and eigenmodes for a range of central link strengths $J_c$. The results are shown in Fig.~\ref{fig:apx:opposed_diamond_tuning}. For $J_c=0$, we recover the spectrum of the four-mode AB-loop defined by the diamond perimeter, with all eigenmodes completely delocalized. 
As discussed in the main text, the introduction of a coupling $J_c>0$ along the central link couples the loop modes of opposite momenta $\tilde{a}_0 \leftrightarrow \tilde{a}_2$ and $\tilde{a}_1 \leftrightarrow \tilde{a}_{-1}$. These become degenerate for $\Phi = 0$ and $\Phi = \pi$, respectively. Consequently, a spectral gap opens at these fluxes. The associated (anti)symmetric superpositions of momentum states fully localize either in the apex or the central resonators. 

\section{Analysis of the $W_5$ wheel graph}\label{sec:apx:W5}
In this section, we include further analysis of the $W_5$ wheel graph network, shown in Fig.~\ref{mpq:fig:wheel5-spectra} and~\ref{mpq:fig:wheel5-transport}. We map the dynamics of the opposite-chirality wheel graph in Fig.~\ref{mpq:fig:wheel5-spectra}a, top, onto an effective three-mode AB loop. This network is dubbed  $\uparrow\downarrow,\boxtimes$. First, we note that the $W_5$ network can be seen as a fusion of two vorticity-aligned diamond networks. This suggests us to express the wheel graph's Hamiltonian matrix $\mathbf{J}_{\uparrow\downarrow\boxtimes}$ in the basis $\{ \tilde{a}_{2-4}, \tilde{a}_{3-5}, \tilde{a}_{2+4}, \tilde{a}_{3+5}, a_1 \}$, where $\tilde{a}_{j\pm k} = (a_j \pm k) / \sqrt{2}$ are symmetric and antisymmetric superpositions of the perimeter resonators, to find
\begin{align}
    \mathbf{J}_{\uparrow\downarrow,\boxtimes}' &= \begin{bmatrix}
    0   & 0     & 0                         & 0                   & 0 \\
    0   & 0     & 0                         & 0                   & 0 \\
    0   & 0     & 0                         & 2 J                 & J_s\sqrt{2}e^{-\iu \Phi} \\
    0   & 0     & 2 J                       & 0                   & J_s\sqrt{2} e^{\iu \Phi} \\
    0   & 0     & J_s\sqrt{2}e^{\iu \Phi}  & J_s \sqrt{2} e^{-\iu \Phi} & 0 \\
    \end{bmatrix} \nonumber \\
    &= \begin{bmatrix}
        0 &   & \\
          & 0 & \\
          &   & \mathbf{J}_3'
    \end{bmatrix}.
    \label{eq:apx:aligned_wheel5_ham_mapped}
\end{align}
Similar to the apex mode in the diamond, the two antisymmetric perimeter modes $\tilde{a}_{2-4}, \tilde{a}_{3-5}$ are decoupled. The remaining modes evolve under the Hamiltonian $\mathbf{J}_3'$ of a three-mode AB loop with couplings $J_a \to 2 J_c$, $J_b \to J_s \sqrt{2}$, pierced by a net flux of $\Phi \to 2\Phi$ that is distributed evenly over the last two links.

\section{Independent prediction of $W_5$ response}
\label{sec:apx:w5_response_calcs}
In the homogeneous $W_5$ network, we predict the response to coherent driving, as shown in Fig.~\ref{mpq:fig:wheel5-transport}a, from the network's susceptibility matrix
\begin{align}
    \mathbf{\chi}_{\vec{a}}(\Delta_\text{m}) = \iu (\Delta_\text{m} \mathbb{1} - \mathbf{J} + \iu \mathbf{\Gamma}/2)^{-1}. \label{eq:apx:susceptibility_mat_def}
\end{align}
Each resonator $j$ can be driven individually, through radiation pressure, by modulating the intensity of a weak drive laser with a coherent tone at a frequency $\omega_{\text{d},j}$ close the mechanical resonance frequency $\Omega_j$. 
In Eq.~\eqref{eq:apx:susceptibility_mat_def}, $\Delta_\text{m} = \omega_{\text{d},j} - \Omega_j$ is the global detuning of all driving tones, while 
\begin{equation}
    \mathbf{J} = \begin{bmatrix}
        0   & J_s           & J_s           & J_s           & J_s \\
        J_s & 0             & J e^{-\iu\Phi}& 0             & J e^{\iu\Phi} \\
        J_s & J e^{\iu\Phi} & 0             & J e^{-\iu\Phi}& 0             \\
        J_s & 0             & J e^{\iu\Phi} & 0             & J e^{-\iu\Phi}\\
        J_s & J e^{-\iu\Phi}& 0             & J e^{\iu\Phi} & 0             
    \end{bmatrix}
\end{equation}
is the network's Hamiltonian matrix and $\mathbf{\Gamma} = \diag(\gamma_1, \dots, \gamma_N)$ encodes the mechanical dissipation rates $\gamma_j$, as listed in Table~\ref{tab:device_params}. 
In our experiment, the drive laser is resonant with the optical cavity.

The vector $\vec{\alpha}$ of the resonators' complex response amplitudes $\alpha_j$, expressed in units of the zero-point motion $\xzpfi{j}$ and in a frame that rotates along with each resonator $j$, is then given by
\begin{align}
    \vec{\alpha} = \mathbf{\chi}_{\vec{a}}(\Delta_\text{m}) \vec{f}.
\end{align}
Each element of $\vec{f}$ encodes the complex amplitude 
\begin{align}
    f_j = \iu e^{\iu \phi_{\text{d},j}} c_{\text{d},j} \frac{g_{0,j} \bar{n}_d}{2}
\end{align}
of the tone driving resonator $j$ with modulation depth $c_{\text{d},j}$, phase offset $\phi_{\text{d},j}$ and corresponding optomechanical coupling rate $g_{0,j}$, while $\bar{n}_d$ denotes the average cavity photon population contributed by the weak, resonant drive laser. 

Drive and response amplitudes are normalised such that
\begin{align}
    \alpha_j = f_j/(\gamma_j/2 - \iu \Delta_m) \label{eq:apx:single_res_response}
\end{align}
when all resonators are uncoupled ($\mathbf{J} = \mathbf{0}$). 
In practice, we establish the value of the proportionality constant $d_j = g_{0,j} \bar{n}_d/2$ between $|f_j|$ and $c_{\text{d},j}$ in a calibration experiment where, for each resonator $j$ individually, we measure $\alpha_j$ as a function of $\Delta_\text{m}$ for fixed $c_{\text{d},j}$. We then fit Eq.~\eqref{eq:apx:single_res_response} to the results to extract $d_j$. For the experiments shown in Fig.~\ref{mpq:fig:wheel5-transport}a, we used driving strengths $f_1/(2\pi) = 1.91$~MHz (left) and $f_2/(2\pi) = 2.07$~MHz (middle and right).

\section{Topology of the $W_5$ lattice}\label{sec:apx:w5_topology}
\begin{figure}[tb]
		\centering
		\includegraphics{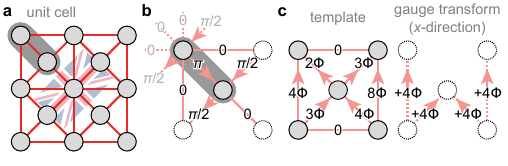}
		\caption{\textbf{Union Jack (UJ) lattice}. 
        \subidc{a} Repeating $W_5$ across the plane forms a lattice resembling the Union Jack (shown in background), known also as the tetrakis square tiling. The UJ lattice's primitive unit cell comprises two resonators (dark grey). 
        \subidc{b} For an infinite, two-dimensional UJ lattice pierced by a homogeneous flux $\Phi = \pi/2$ per plaquette, the magnetic unit cell (dark grey) equals the lattice unit cell. Solid lines indicate the links we associate this unit cell but include couplings to neighboring cells (dotted circles). Dashed lines indicate links associated with neighboring cells. Link phases are shown with direction (arrows). Note that link phase $\pi$ is equivalent to $-\pi$.
        \subidc{c} Template cell to construct a UJ lattice pierced by an arbitrary, homogeneous flux $\Phi$. Link phases are expressed in a Landau gauge and vary only in the $x$-direction. The indicated gauge transform constructs each next cell to the right, while cells are copied as-is to construct adjacent cells in the $y$-direction.
        \label{fig:apx:union_jack_lattice}
        }
\end{figure}

In the main text, we have explored $W_5$ graphs threaded by synthetic fluxes. By concatenating these $W_5$ building blocks, one can construct the Union Jack (UJ) lattice, where each $W_5$ graph forms the fundamental unit (Fig.~\ref{fig:apx:union_jack_lattice}a). This lattice is generated by subdividing the squares of a standard square lattice into triangles, with the diagonals and edges naturally forming the connections of the $W_5$ structure. Within the $W_5$ lattice, the sites can be distinguished into two sublattices based on their roles and connectivity. The central or ``hub'' sites of the $W_5$ graphs form one sublattice, and link directly to the four surrounding peripheral or ``rim'' sites. The peripheral sites, which belong to the second sublattice, form a closed loop and are shared between neighboring $W_5$ units.

\begin{figure*}[tb]
		\centering
	\includegraphics[width=\textwidth,keepaspectratio]{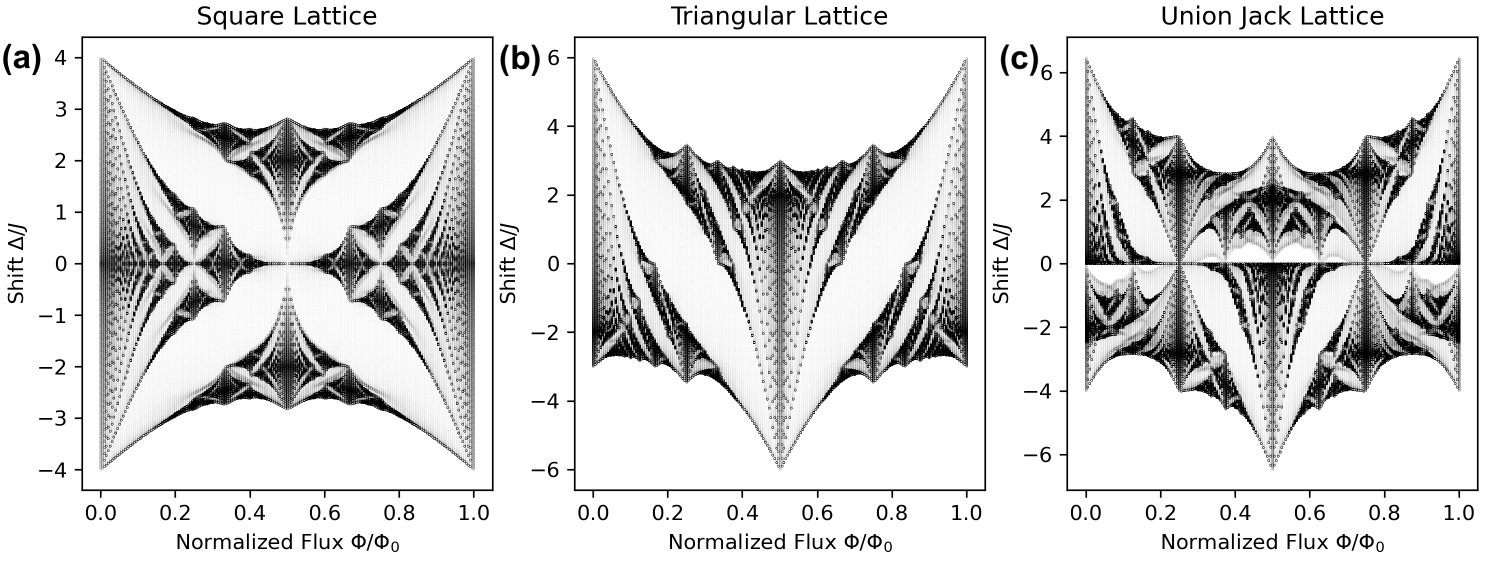}  
		\caption{Hofstadter butterfly spectra for the square, triangular, and Union Jack lattices. Each panel shows the fractal energy spectrum as a function of the magnetic flux per plaquette, \(\Phi = 2\pi \frac{p}{q}\). \subidc{a} The square lattice exhibits a symmetric and well-known fractal pattern. \subidc{b} The triangular lattice features a more intricate spectrum with distorted subbands due to its denser connectivity. \subidc{c} The Union Jack lattice combines characteristics of the square and triangular lattices, resulting in a ``hybrid'' spectrum with unique gap patterns and subband structures. All figures are normalized to a flux quantum $\Phi_0$, defined by the plaquette area of each lattice. For the square lattice, $\Phi_0 = 2\pi$. For the triangular lattice, the smaller triangular plaquettes give $\Phi_0 = 8\pi/\sqrt{3}$, while the UJ lattice $\Phi_0 = 8\pi$. This ensures consistent flux normalization across lattices.}
    \label{fig:hofstadter_butterflies}
\end{figure*}

In the simulations, the Union Jack lattice is constructed by first defining a basis with two sites: one site $a$ at the origin $(0, 0)$ and another site $b$ at $(\ell/2, \ell/2)$ –– the site to become a hub. The primitive vectors of the lattice are then chosen as $\vec{\ell}_1 = (\ell, 0)$ and $\vec{\ell}_2 = (0, \ell)$, which define a square unit cell of side length $\ell$. We label each unit cell by its position indices $j$ and $k$ along the lattice vectors $j \vec{\ell}_1 + k \vec{\ell}_2$. The standard nearest-neighbor connections between the sites $a_{j,k}$ at the corners of the squares form the perimetral structure of the lattice. Diagonal connections are then added between each of these corner sites and the central site $b_{j,k}$ in the same and neighboring unit cells, creating the characteristic structure of the UJ lattice.

\subsection{Hofstadter butterfly spectra}
The Hofstadter butterfly spectrum describes the energy spectrum of excitations on a 2D periodic lattice subject to a synthetic flux per plaquette, $\Phi = 2\pi p/q$, where $p$ and $q$ are coprime integers. The spectrum, plotted as a function of $\Phi$, reveals a fractal pattern of gaps and bands resulting from the interplay between the geometric length, $l_g \sim \ell$, set by the lattice spacing, and the magnetic length, $l_\Phi \sim 1/\sqrt{\Phi}$, which characterizes the spatial extent of the synthetic flux's influence. In the square lattice, when $\Phi$ is rational, $l_g$ and $l_\Phi$ commensurate, splitting each energy band into $q$ magnetic subbands and producing the fractal Hofstadter butterfly~\cite{Hofstadter1976}, Fig.~\ref{fig:hofstadter_butterflies}\subidc{a}.

For the triangular lattice, the Hofstadter butterfly appears more intricate. Each energy band splits into $q$ magnetic subbands when the flux per plaquette is $\Phi = 2\pi p/q$~\cite{Hasegawa1989,Du2018}, Fig.~\ref{fig:hofstadter_butterflies}\subidc{b}. The shorter path connectivity of the triangular lattice distorts the fractal symmetry and modifies the gap structure compared to the square lattice, resulting in a richer but less symmetric spectrum. For the Union Jack lattice, shown in Fig.~\ref{fig:hofstadter_butterflies}\subidc{c}. the Hofstadter butterfly spectrum combines features of both the square and triangular lattices. The hybrid connectivity of the Union Jack lattice creates a unique energy spectrum where the fractal structure of the Hofstadter butterfly still emerges but with distinct gap patterns and subband structures characteristic of its geometry. This spectrum reflects the influence of both square-like perimetral bonds and the central hub connections, highlighting the interplay of the lattice’s hybrid nature with the magnetic flux.

\subsection{Band structure of the UJ lattice with $\Phi=\pi/2$}
For a planar UJ lattice with translational invariance in both directions,
where each triangular plaquette is pierced by a homogeneous flux $\Phi = \pi/2$, the magnetic unit cell (Fig.~\ref{fig:apx:union_jack_lattice}b) is equal to the lattice unit cell. The Hamiltonian $H$ that governs the lattice is specified by the matrix elements
\begin{align}
\begin{aligned}
    \braket{a_{j,k} | H | a_{j+1,k}} &= J, 
    &\braket{a_{j,k} | H | a_{j,k+1}} &= J, \\
    \braket{a_{j,k} | H | b_{j,k}}   &= -J, 
    &\braket{b_{j,k} | H | a_{j+1,k+1}} &= J, \\
    \braket{b_{j,k} | H | a_{j,k+1}} &= -\iu J, 
    &\braket{a_{j+1,k} | H | b_{j,k}} &= -\iu J,
\end{aligned}
\end{align}
and their Hermitian conjugates. These elements encode the couplings within and between unit cells as shown in Fig.~\ref{fig:apx:union_jack_lattice}b by solid links, where $J$ is the coupling strength.

To calculate the band structure of this infinite, two-dimensional lattice, we take the dynamical matrix of the magnetic unit cell and apply the Bloch theorem for periodic boundary conditions,
\begin{align}
    \ket{(a,b)_{j+n,k+m}} &= e^{\iu{(n k_x + m k_y)}} \ket{(a,b)_{j,k}}, \label{eq:apx:UJ_Bloch_BC}
\end{align}
for wavevectors $k_x$ and $k_y$ in the $x$ and $y$-directions, respectively.
The matrix elements of $H$ that drive rim site $a_{j,k}$ correspond to all links that connect to the $a$-site in Fig.~\ref{fig:apx:union_jack_lattice}b (solid and dashed), and are given by
\begin{align}
\begin{aligned}
    \braket{a_{j,k} | H | a_{j+1,k}} &= J, 
    &\braket{a_{j,k} | H | a_{j,k+1}} &= J, \\
    \braket{a_{j,k} | H | a_{j-1,k}} &= J, 
    &\braket{a_{j,k} | H | a_{j,k-1}} &= J, \\
    \braket{a_{j,k} | H | b_{j,k}} &= -J, 
    &\braket{a_{j,k} | H | b_{j-1,k}} &= -\iu J, \\
    \braket{a_{j,k} | H | b_{j-1,k-1}} &= J,
    &\braket{a_{j,k} | H | b_{j,k-1}} &= \iu J.
\end{aligned} \label{eq:apx:UJ_Bloch_a_drives}
\end{align}
Likewise, the matrix elements that drive hub site $b_{j,k}$ are given by
\begin{align}
\begin{aligned}
    \braket{b_{j,k} | H | a_{j,k}} &= -J, 
    &\braket{b_{j,k} | H | a_{j+1,k}} &= \iu J, \\
    \braket{b_{j,k} | H | a_{j,k+1}} &= -\iu J, 
    &\braket{b_{j,k} | H | a_{j+1,k+1}} &= J.
\end{aligned} \label{eq:apx:UJ_Bloch_b_drives}
\end{align}

By combining Eqs.~\eqref{eq:apx:UJ_Bloch_a_drives} and~\eqref{eq:apx:UJ_Bloch_b_drives} with boundary condition~\eqref{eq:apx:UJ_Bloch_BC}, we arrive at the $2\times 2$ Bloch Hamiltonian $H_B$ for the unit cell sites $a$ and $b$, specified by matrix elements
\begin{align}
\begin{aligned}
    \bra{a} H_B \ket{a} &= J \left( e^{-\iu k_x} + e^{\iu k_x} + e^{-\iu k_y} + e^{\iu k_y} \right) \nonumber \\
    &= 2J \left( \cos(k_x) + \cos(k_y) \right), \\
    \bra{a} H_B \ket{b} &= J \left( e^{-\iu (k_x + k_y)} - 1 - \iu e^{-\iu k_x} + \iu e^{-\iu k_y}  \right), \\
    \bra{b} H_B \ket{a} &= J \left( e^{\iu (k_x + k_y)} - 1 + \iu e^{\iu k_x} -\iu e^{\iu k_y} \right), \\
    \bra{b} H_B \ket{b} &= 0.
\end{aligned}
\end{align}
The band structure of $H_B$ is given by its two eigenvalues
\begin{align}
    \frac{\epsilon_\pm}{J} = \cos(k_x) + \cos(k_y) \pm \sqrt{4+(\cos(k_x) - \cos(k_y))^2},
\end{align}
which are plotted in Fig.~\ref{mpq:fig:wheel5-transport}c.

\subsection{Chern numbers for flux $\Phi=\pi/2$}
We study the integer quantum Hall effect, a well-known topological phase of matter, in this UJ lattice. Topological phases exhibit properties that are robust against local perturbations, with quantized conductance values tied to global invariants in condensed-matter systems. In this case, the quantized Hall conductance is linked to the topology of the system's electronic bands, described by the Chern number. Similar concepts extend to topological bosonics, e.g. topological photonics, where the Chern number governs robust light transport, enabling unidirectional edge states immune to backscattering~\cite{Ozawa2019topological}

Starting from the Bloch eigenstates $|u_\pm(\vec{k})\rangle$ corresponding to the two eigenvalues $\epsilon_\pm$ of $H_B$, the Berry curvature $\Omega_\pm$ is calculated for each band as
\begin{equation}\label{eq:BC}
\Omega_\pm(\vec{k}) = 2\text{Im} \left\langle \partial_{k_x} u_\pm(\vec{k}) \middle| \partial_{k_y} u_\pm(\vec{k}) \right\rangle,
\end{equation}
having only one component in 2D. The Berry curvature acts as a momentum-space curvature, and its integral reveals the topological nature of the band. The net flux of Berry curvature over the Brillouin zone (BZ) gives the Chern number:
\begin{equation}
C_\pm = \frac{1}{2\pi} \int_\text{BZ} \mathrm{d}\vec{k}\hspace{2mm} \Omega_\pm(\vec{k}).
\end{equation}
This invariant reflects the global structure of the band in a manner analogous to the Gauss-Bonnet theorem,  where the integral of Gaussian curvature encodes a topological property (the Euler characteristic) of a surface~\cite{do2016differential}.  We calculate the Chern numbers for the UJ lattice symbolically, using \textsc{Mathematica} software, and find
\begin{equation}
    C_\pm = \pm 1.
\end{equation}
This result indicates that the UJ lattice hosts topologically nontrivial phases, with $C_+ = 1$ and $C_- = -1$ corresponding to chiral edge states propagating in opposite directions, as per the bulk-boundary correspondence.

\subsection{Direct retrieval of edge states}
A ribbon geometry imposes translational symmetry along the $y$-direction (infinite) while keeping the network finite in the $x$-direction to form a ribbon of fixed width $N$. This setup isolates edge states, allowing the study of chiral modes induced by the flux $\Phi$.

A natural gauge choice for the ribbon is the Landau gauge, where link phases vary only in the $x$-direction. In Fig.~\ref{fig:apx:union_jack_lattice}c, we show how the UJ lattice can be constructed with arbitrary flux $\Phi$ in this gauge. The ribbon consists of a single row of $N$ primitive unit cells, labeled by their index $j$ along the $x$-direction, and comprises the sites $a_j$ and $b_j$. To ensure termination on a rim site on both ends, we remove the final hub site $b_N$.

To calculate the band structure of the infinite ribbon (as plotted in Fig.~\ref{mpq:fig:wheel5-transport}d) and the corresponding eigenstates (as plotted in Fig.~\ref{mpq:fig:wheel5-transport}e), we construct the Bloch Hamiltonian $H_B$ in a similar fashion as before. We find that it is specified by the matrix elements
\begin{align}
\begin{aligned}
    \bra{a_j} H_B \ket{a_j} &= 2 J \cos(k_y-4 j \Phi), \\
    \bra{a_j} H_B \ket{b_j} &= J \left( e^{2 \iu \Phi} + e^{-\iu k_y} e^{-\iu (3+4j) \Phi} \right), \\
    \bra{b_j} H_B \ket{a_j} &= J \left( e^{-2\iu\Phi} + e^{\iu k_y} e^{\iu (3+4j)\Phi} \right), 
\end{aligned}
\end{align}
that couple site within a cell $j$, and the inter-cell matrix elements
\begin{align}
\begin{aligned}
    \bra{a_j} H_B \ket{a_{j+1}} &= J, \\
    \bra{a_{j+1}} H_B \ket{a_j} &= J, \\
    \bra{b_j} H_B \ket{a_{j+1}} &= J \left( e^{-3 \iu \Phi} + e^{\iu k_y} e^{\iu (4+4j)\Phi} \right), \\
    \bra{a_{j+1}} H_B \ket{b_j} &= J \left( e^{3\iu \Phi} + e^{-\iu k_y} e^{-\iu (4+4j)\Phi}\right),
\end{aligned}
\end{align}
while all other elements are zero. Again, the eigenvalues of $H_B$ determine the dispersion of the ribbon, while the corresponding eigenvectors determine its wave states.


%

\end{document}